\documentclass[twocolumn,showpacs,fleqn,nobibnotes,prd]{revtex4}
\usepackage{amsmath}
\usepackage{graphicx}
\usepackage{float}

\usepackage{subfigure}

\def\lsim{\raise0.3ex\hbox{$<$\kern-0.75em\raise-1.1ex\hbox{$\sim$}}}
\def\gsim{\raise0.3ex\hbox{$>$\kern-0.75em\raise-1.1ex\hbox{$\sim$}}}

\newcommand{\rr}{\mbox{\boldmath $r$}}

\newcommand{\rb}{\mbox{\boldmath $b$}}

\newcommand{\be}{\begin{equation}}
\newcommand{\ee}{\end{equation}}

\begin{document}
\title{High density effects in  ultrahigh neutrino interactions}

\author{V.P. Gon\c{c}alves and D.R. Gratieri}

\affiliation{Instituto de F\'{\i}sica e Matem\'atica, Universidade Federal de
Pelotas\\
Caixa Postal 354, CEP 96010-900, Pelotas, RS, Brazil.
}

\begin{abstract}

The high parton density present at  high energies and large nuclei is expected to modify the lepton - hadron cross section and the associated observables. In this paper we analyse the impact of the high density effects in the average inelasticity and the neutrino - nucleus cross section at ultra high energies. We compare the predictions associated to the linear DGLAP dynamics with those from the  Color Glass Condensate formalism, which includes non-linear effects. Our results demonstrated that the non-linear effects reduce the average inelasticity and that the predictions of the distinct approaches for the neutrino - nucleus cross section at ultra-high energies are similar.

\end{abstract}

\maketitle

\section{Introduction}

The recent detection of  ultra-high energies (UHE) neutrinos by the IceCube Neutrino Observatory \cite{IceCube} starts a new era in the neutrino physics. These and forthcoming data may shed light on the observation of air showers events with energies in excess of $10^{11}$ GeV, reveal aspects of new physics as well as of the QCD dynamics at high energies (For recent reviews see, e.g., Ref. \cite{reviews_neutrinos}). 

One of the main ingredients for estimating event rates at IceCube is the neutrino - hadron cross section ($\sigma_{\nu h}$).  The behaviour of $\sigma_{\nu h}$ at high energies, which 
provides a probe of Quantum Chromodynamics (QCD) in the kinematic region of very small values of Bjorken-$x$, not explored by the HERA measurements \cite{hera},   has been discussed by several authors  considering different approaches \cite{varios,HEPP}. 
Currently there is a large uncertainty in the predictions, directly associated with the uncertainty present in the treatment of the QCD dynamics at high energies.
Theoretically, at high energies (small Bjorken-$x$)  one
expects the transition of the regime described by the DGLAP
dynamics \cite{dglap}, where only the parton emissions are considered, to a new
regime where the physical process of recombination of partons becomes
important in the parton cascade and the evolution is given by a
non-linear evolution equation \cite{cgc}.  This regime is characterized by the
limitation on the maximum phase-space parton density that can be
reached in the hadron wavefunction (parton saturation), with the
transition being specified  by a typical scale, which is energy and atomic number
dependent and is called saturation scale $Q_{\mathrm{s}}$ .
Moreover,  the growth of the parton distribution is expected to saturate, forming a  Color Glass Condensate (CGC) (For recent reviews see Ref. \cite{hdqcd}). 
Although a large amount of data at HERA, RHIC and LHC data can be described by the CGC formalism, the determination  of the magnitude and kinematical region where the saturation effects cannot be disregarded still is an open question.

In Ref. \cite{HEPP} we examined to what extent the cross section is sensitive to the presence of new dynamical effects in the QCD evolution. 
We have compared the more recent predictions based on the  NLO DGLAP evolution equation with those from the Color Glass Condensate physics. Our results demonstrated that the current theoretical uncertainty for the neutrino-nucleon cross section reaches a factor three for neutrino energies around  $10^{11}$ GeV and increases to 5.5 for   $E_{\nu} = 10^{13}$ GeV.  In this paper we  complement the study performed in Ref. \cite{HEPP} in two points. The first one, is the study of the average inelasticity at high energies considering different approaches for the neutrino - nucleon cross section. This analysis is motivated by the results presented in Ref. \cite{zas}, where the authors  demonstrated that the inelasticity, which is the fraction of the neutrino energy which flows to the hadronic part of the interaction in the laboratory frame, is strongly dependent on the high energy behaviour of the cross section. The second point is the study of neutrino -  nucleus interactions at high energies. In this case the high parton density present in the nucleus is expected to modify the parton distributions and amplify the  saturation effects. We update previous DGLAP studies \cite{zas_nuclear} considering here the EPS09 parametrization  \cite{eps} for the nuclear parton distributions and estimate the neutrino - nucleus cross section considering the saturation effects and different models to describe the dipole - nucleus interaction. As in \cite{HEPP}, our goal in this study  is to estimate  the theoretical uncertainty in the neutrino - nucleus cross section  at ultra high energies by considering very distinct models.

The paper is organized as follows. 
In the next section  we present a brief review of the formalism for the treatment of neutrino  deep inelastic scattering (DIS) in terms of the  QCD improved parton model and using the color dipole formalism, which is useful to include the non-linear effects. In Section \ref{ymed} we present the different models used to calculate the neutrino - nucleon cross section and present our predictions for the average inelasticity. 
The neutrino - nucleus cross section is discussed in Section \ref{sec_nuclear} considering the linear DGLAP formalism and the EPS09 parametrization as well as different models for the dipole - nucleus cross section. Moreover, we present our predictions for the nuclear effects in the total cross section. Finally, in Section \ref{conc} we present a summary of our main conclusions.

\section{The neutrino - hadron cross section at high energies}
\label{cross}

Deep inelastic neutrino scattering is described in terms of charged current (CC) and neutral current (NC) interactions, which proceed through $W^{\pm}$ and $Z^0$  exchanges, respectively.    As the neutral current (NC) interactions are subdominant,  we will consider in what follows, for simplicity, only charged current (CC) interactions. The total neutrino - hadron cross section is given by \cite{book}
\begin{eqnarray}
\sigma_{\nu h}^{CC} (E_\nu) = \int_{Q^2_{min}}^s dQ^2 \int_{Q^2/s}^{1} dx \frac{1}{x s} 
\frac{\partial^2 \sigma^{CC}}{\partial x \partial y}\,\,,
\label{total}
\end{eqnarray}
where $E_{\nu}$ is the neutrino energy, $s = 2 ME_{\nu}$ with $M$ the hadron mass, $y = Q^2/(xs)$ and $Q^2_{min}$ is the minimum value of $Q^2$ which is introduced in order to stay in the deep inelastic region. In what follows we assume $Q^2_{min} = 1$ GeV$^2$. Moreover, the differential cross section is given by \cite{book}
\begin{widetext}
\begin{eqnarray} 
\frac{\partial^2 \sigma_{\nu h}^{CC}}{\partial x \partial y} = \frac{G_F^2 M E_{\nu}}{\pi} \left(\frac{M_W^2}{M_W^2 + Q^2}\right)^2 \left[\frac{1+(1-y)^2}{2} \, F_{2,CC}^h(x,Q^2) - \frac{y^2}{2}F_{L,CC}^h(x,Q^2)+ y (1-\frac{y}{2})xF_{3,CC}^h(x,Q^2)\right]\,\,,
\label{difcross}
\end{eqnarray}
\end{widetext}
where $h = p$ or $A$, with $A$ the atomic number, $G_F$ is the Fermi constant and $M_W$ denotes the mass of the charged gauge boson. 
The calculation of $\sigma_{\nu h}$ involves  integrations over $x$ and $Q^2$, with the integral being dominated by  the interaction with partons of lower $x$ and  $Q^2$ values of the order of the electroweak boson mass squared.

In the QCD improved parton model the structure functions $F_2,  \,F_L$ and $F_3$ are calculated in terms of quark and gluon distribution functions. In this case the neutrino - hadron cross section for charged current interactions on an isoscalar target is given by (See, e.g. Ref. \cite{book}):
\begin{widetext}
\begin{eqnarray} 
\frac{\partial^2 \sigma_{\nu h}^{CC}}{\partial x \partial y} = \frac{2 G_F^2 M E_{\nu}}{\pi} \left(\frac{M_W^2}{M_W^2 + Q^2}\right)^2 \left[
xq_h(x,Q^2) + x\bar{q}_h(x,Q^2)(1-y)^2\right]
\label{difcross_partonic}
\end{eqnarray}
\end{widetext}
with the quark and antiquark densities given by $q = (d+u)/2 + s +b$ and $\bar{q} = (\bar{d}+\bar{u})/2 + c+ t$. In what follows, we consider that the evolution of the parton distributions is given by the DGLAP equations and use the CT10 parametrization \cite{cteq}  in our calculations of the neutrino - nucleon cross section. In the nuclear case, we also consider that the parton distributions are modified by nuclear effects as described by the EPS09 parametrization \cite{eps}. The resulting predictions  for the neutrino - hadron cross section characterize the linear QCD dynamics.

 In order to  introduce non-linear (saturation) effects in the QCD dynamics, we will describe  the structure functions considering the color dipole approach in which the DIS to low $x$ can be viewed as a result of the interaction of a color $q\bar{q}$ dipole which the gauge boson fluctuates \cite{nik}.  In this approach the $F_2^{CC}$ structure function is expressed in terms of the transverse and longitudinal structure functions, $F_2^{CC}=F_T^{CC} + F_L^{CC}$ which are given by 
\begin{eqnarray}
&\,& F_{T,L}^{CC}(x,Q^2)  = \nonumber \\
  &\,& \frac{Q^2}{4\pi^2}  \int_0^1 dz \int d^2  \rr |\Psi^{W}_{T,L}(\rr,z,Q^2)|^2 \sigma^{dh}(\rr,x)\,\,
\label{funcs}
\end{eqnarray} 
where $r$ denotes the transverse size of the dipole, $z$ is the longitudinal momentum fraction carried by a quark and  $\Psi^{W}_{T,L}$ are proportional to the wave functions of the virtual charged gauge boson corresponding to their transverse or longitudinal polarizations.  Furthermore, $\sigma^{dh}$ describes the interaction of the  color dipole with the target. 
In the Color Glass Condensate formalism the dipole - target cross section 
$\sigma^{dh}$ can be computed in the eikonal approximation,
resulting
\begin{eqnarray}
\sigma^{dh} (x,\rr)=2 \int d^2 \rb \, {\cal{N}}^h(x,\rr,\rb)\,\,,
\label{dipnuc}
\end{eqnarray}
where ${\cal{N}}^h$ is the forward dipole-target scattering amplitude
for a given impact parameter $\rb$  which encodes all the
information about the hadronic scattering, and thus about the
non-linear and quantum effects in the hadron wave function. 
It is useful to assume that the impact parameter dependence of $\cal{N}$
can be factorized as ${\cal{N}}(x,\rr,\rb) = {\cal{N}}(x,\rr)
S(\rb)$, so that $\sigma^{dp}(x,\rr) = {\sigma_0}
\,{\cal{N}}(x,\rr)$, with $\sigma_0$ being   a free parameter
related to the non-perturbative QCD physics. The Balitsky-JIMWLK
hierarchy \cite{cgc,bk}  describes the energy evolution of the dipole-target
scattering amplitude ${\cal{N}}(x,\rr)$.
In the mean field approximation, the first equation of this  hierarchy decouples and boils down to the Balitsky-Kovchegov (BK) equation \cite{bk}.

In the last years 
the next-to-leading order corrections to the  BK equation were calculated  
\cite{kovwei1,javier_kov,balnlo} through the ressumation of $\alpha_s N_f$ contributions to 
all orders, where $N_f$ is the number of flavors. Such calculation allows one to estimate 
the soft gluon emission and running coupling corrections to the evolution kernel.
The authors have verified that  the dominant contributions come from the running 
coupling corrections, which allow us to  determine the scale of the running coupling in the 
kernel. The solution of the improved BK equation was studied in detail in Refs. 
\cite{javier_kov,javier_prl}. Basically, one has that the running of the coupling reduces 
the speed of the evolution to values compatible with experimental data, with the geometric 
scaling regime being reached only at ultra-high energies. In \cite{bkrunning} a global 
analysis of the small $x$ data for the proton structure function using the improved BK 
equation was performed  (See also Ref. \cite{weigert}). In contrast to the  BK  equation 
at leading logarithmic $\alpha_s \ln (1/x)$ approximation, which  fails to describe the HERA 
data, the inclusion of running coupling effects in the evolution renders the BK equation 
compatible with them (See also \cite{vic_joao,alba_marquet,vicmagane}).

In Ref. \cite{HEPP} we considered different models for the forward dipole - proton scattering amplitude. In particular, the solution of the running coupling Balitsky - Kovchegov equation \cite{bkrunning}, which is currently the  state-of-art for the description of saturation effects, was used as input in our calculations. This solution  will also be used in our calculations of the average inelasticity in the next Section. On the other hand, in the calculations of the neutrino - nucleus cross section  it is necessary to specify the forward dipole - nucleus scattering amplitude, ${\cal{N}}^A$, which still is an open question and is usually described by phenomenological models. In Section \ref{sec_nuclear} we will present some possible models for this quantity and analyse its implications in $\sigma^{\nu A}$.

\section{The average inelasticity}
\label{ymed}
The inelasticity $y$ is the fraction of the neutrino energy  transferred to the hadronic target in the laboratory frame. It is responsible for the relative sizes of the electromagnetic and hadronic showers induced in a charged current neutrino interaction. Its precise determination is fundamental in order to extract the neutrino energy in high energy neutrino telescopes from the detected muon or hadronic showers. In what follows we estimate the average value of the inelasticity, $\langle y \rangle$, which is given by
\begin{eqnarray}
\langle y \rangle = \frac{\int^{1}_{0} dy~ y ~\frac{d\sigma}{dy}}{\int^{1}_{0} dy~ \frac{d\sigma}{dy}}
\end{eqnarray}
with $d\sigma/dy$ obtained by the integration of the Eq. (\ref{difcross}). In Ref. \cite{zas} the authors demonstrated that the behaviour of $\langle y \rangle$ at large energies is directly related to the $x$ slope of the parton distributions at small-$x$ and $Q^2 \approx M_W^2$. In other words, that the average rapidity is sensitive to the QCD dynamics at high energies, which  determines the high energy behaviour of the cross section.  Our goal is to update this analysis considering the more recent approaches for the QCD dynamics.

\begin{figure}[t]
\includegraphics*[scale=0.3]{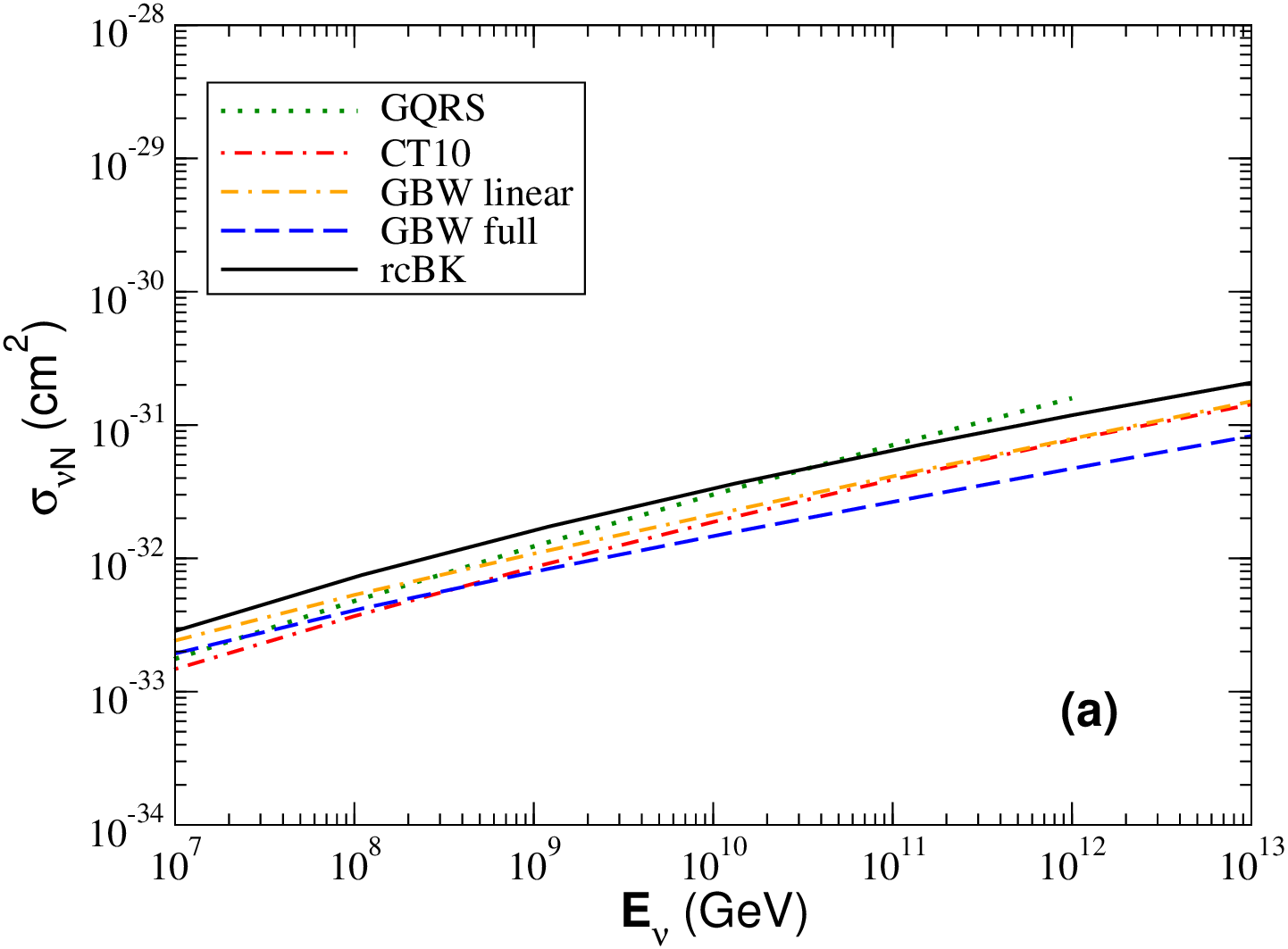}
\includegraphics*[scale=0.3]{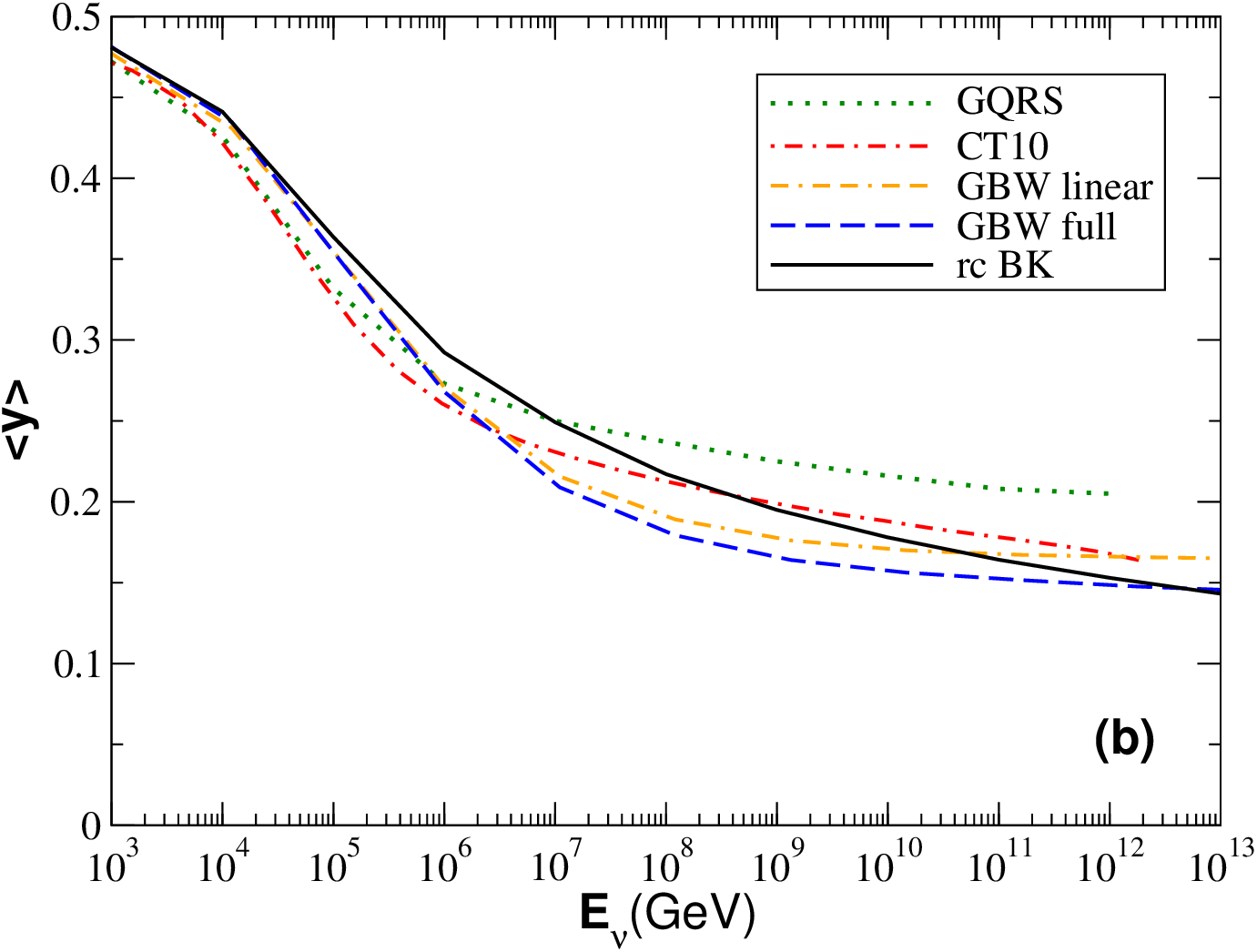}
\caption{ (Color online) {(a)}: Energy dependence of the neutrino-nucleon cross-section predicted by different linear and non-linear models.    {(b)} Energy dependence of  the average inelasticity $<y>$.}
\label{0123}
\end{figure}




In Fig.~\ref{0123} (a) we present our predictions for the neutrino-proton  cross-section considering different models for dynamics, which will be used in the calculations of the average inelasticity. In particular, we present the linear prediction obtained using the CT10 parametrization \cite{cteq} and for comparison the GQRS one \cite{gqrs}, which have been used in several analysis since its proposition. The CT10 prediction has a smother growth with energy in comparison with the GQRS one, which is directly associated to the distinct behaviours at small-$x$ of the gluon distributions present in these calculations. A more detailed comparison with other linear models is presented in Ref. \cite{HEPP}. We also present the predictions obtained using the solution of the rcBK equation and the phenomenological saturation model proposed by Golec - Biernat and Wusthoff in Ref. \cite{gbw} (denoted GBW hereafter), in which the dipole - proton cross section is given by:
\begin{eqnarray}
\sigma^{dp}_{GBW} = \sigma_0 \left[1-\exp(-\frac{r^2Q_{s,p}^2(x)}{4})\right],
\end{eqnarray}
where the saturation scale is given by $Q_{s,p}^2(x)=Q_0^2\left(x_0/x\right)^{\lambda}$, with $Q_0^2 = 1$ GeV$^2$, $x_0 =3\,.\,10^{-4} $ and $\lambda = 0.288$. Our motivation to use this model, which have been updated in several aspects in the last years, is that it allow us easily to obtain its linear limit,  given by  $\sigma^{dp}_{GBW\,linear} = \sigma_0 {r^2Q_{s,p}^2(x)}/{4}$. Consequently, it allows to quantify the contribution 
of the saturation effects in the observable under analysis. In particular, the difference between the linear and full GBW predictions for the neutrino - proton cross section at large energies is a factor $\approx 2$. It is important to emphasize the similarity between the CT10 and GBW linear predictions at high energies. 
As already demonstrated in Ref. \cite{HEPP} the rcBK model predicts  a different normalization and a steeper energy dependence in comparison to the GBW one.

In Fig.~\ref{0123} (b)
 we show our results for the energy dependence of the average inelasticity considering the models discussed above. 
All models agree in the predictions for $\langle y \rangle$ in the kinematical region of low values of the neutrino energy. However,
for energies above $10^4$ GeV, the distribution is shifted to lower values of $y$ due to 
 the $W$ propagator which acts as a cut-off in the integration restricting the values of $Q^2$ to around $M_W^2$  \cite{zas}.
 Moreover, 
 the predictions differ by $\approx 30 \%$ at ultra-high energies, with the GQRS prediction ($\langle y \rangle \approx 0.20$) being the upper bound and the GBW one ($\langle y \rangle \approx 0.15$)   the lower bound. This result confirms that the average inelasticity is sensitive to the QCD dynamics at high energies \cite{zas}.  The distinct gluon distributions  present in the GRQS and CT10 calculations, with the latter being flatter, implies a reduction of  $\langle y \rangle$ at ultra-high energies. Considering the non-linear approaches, the difference between the rcBK and GBW predictions is directly associated to the transition between the linear and saturation regimes, which is delayed in the rcBK approach, which implies the distinct behaviours in the $10^{5} \le E_{\nu} \le 10^{12}$ range. At ultra-high energies, both approaches predict the saturation of  the dipole - nucleon cross section, which implies similar predictions at  $E_{\nu} >  10^{12}$ GeV. Comparing the linear and full GBW predictions, we obtain that the contribution of saturation effects is important for $E_{\nu} >  10^{6}$ GeV, reducing $\langle y \rangle$ in approximately 10$\%$. Finally, the CT10 prediction seems to agree with the GBW-linear  one at   $E_{\nu}\approx 10^{12}$.

\section{Neutrino - nucleus cross section}
\label{sec_nuclear}

 In order to calculate the neutrino - nucleus cross section we need to take into account the nuclear effects in the structure functions (For a recent review see Ref. \cite{kope_review}). 
  However, after more than 30 years of experimental and theoretical studies, a standard picture of nuclear modifications of structure functions and parton densities has not yet emerged \cite{armesto_review,frank_review}. Fixed target DIS measurement on nuclei revealed that the ratio of nuclear to nucleon structure functions (normalized by the atomic mass number) is significantly differently than unity.
In particular, these data demonstrate an intricate behaviour, with the ratio being less than one at large $x$ (the EMC effect) and at small $x$ (shadowing) and larger than one for $x \approx 10^{-1}$ (antishadowing).  
  The existing data were taken at lower energies  and therefore  the perturbative QCD regime ($Q^2  \ge 1$ GeV$^2$) was explored  only for relatively large values of the (Bjorken) $x$  variable ($x > 10^{-2} $). Experimentally, this situation will hopefully change with  a future high energy electron-ion collider (EIC) (For recent reviews see, e.g. \cite{erhic,lhec}),  which is supposed to take data at higher energies and explore the region of small $x$ ($   x < 10^{-2} $) in the perturbative QCD regime.    From the theoretical point of view the theory of nuclear effects in deep inelastic scattering (DIS) is still far from being concluded. The straightforward use of nucleon parton distributions evolved with DGLAP equations and corrected with a nuclear  modification factor  determined by fitting the existing data as in Refs. \cite{eks,hkn,ds,deflorian,eps,cteq_nuclear} is only well justified in the large $Q^2$ region and not too small $x$. However, these approaches do not address the fundamental problem of the origin of the nuclear shadowing and  cannot be extended at small $x$ where we expect to see  new interesting physics related to the non-linear aspects of QCD dynamics  \cite{hdqcd}. In order to obtain an uncertainty band for the neutrino - nucleus cross section we consider two extreme scenarios: (a) the linear one, with the nuclear structure functions described in terms of parton distributions which evolve according the DGLAP equations, and (b) the non-linear one, with the nuclear structure functions described in terms of the dipole - nucleus scattering amplitude, which takes into account the saturation effects. In this case we consider some possible models for ${\cal{N}}^A$ which are found in the literature.

\begin{figure}[t]
\includegraphics*[scale=0.3]{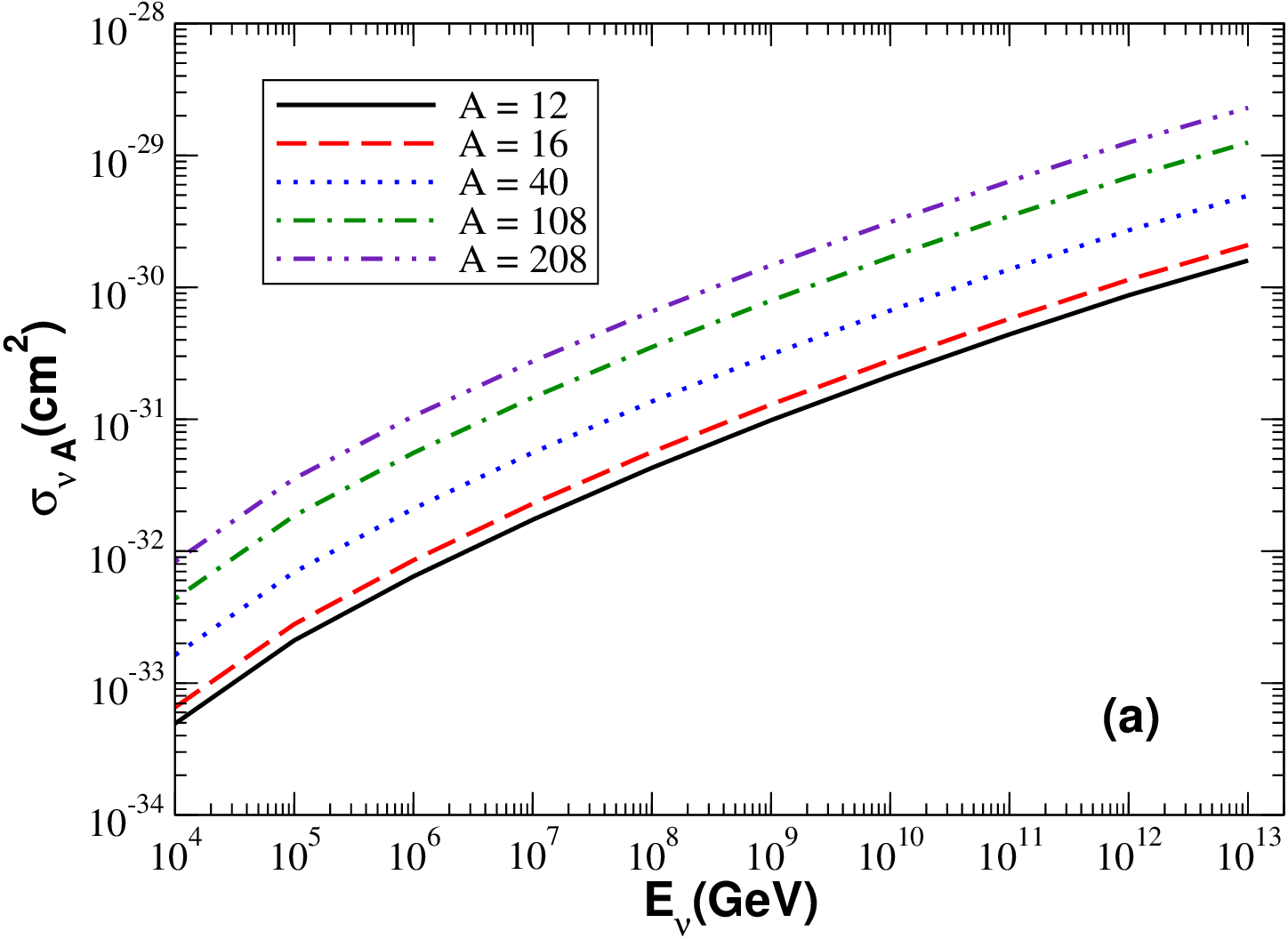}
\includegraphics*[scale=0.3]{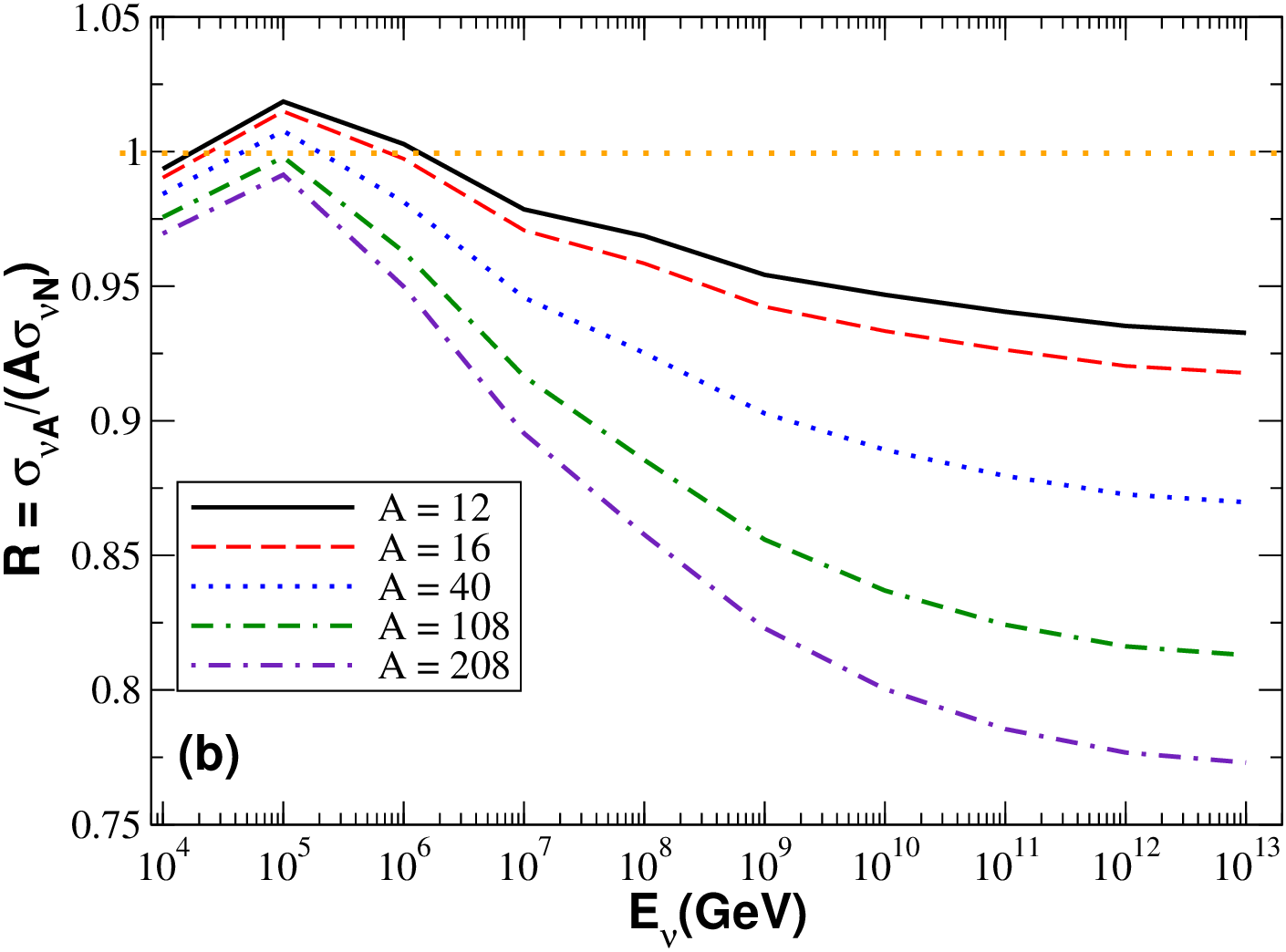}
\caption{(Color online) {(a)} Energy dependence of the neutrino - nucleus cross-section for different nuclei; {(b)}: Energy dependence of the nuclear ratio $R = \sigma_{\nu A} / A\,\sigma_{\nu N}$.}
\label{Ratio-nucle}
\end{figure}

Lets start our analysis considering the linear scenario for the treatment of nuclear effects in the neutrino - nucleus cross section at high energies.     
  We will make use of   the EPS09 parametrization \cite{eps} of the nuclear parton distribution functions (nPDFs), which is  based on a global fit of the nuclear data using the DGLAP evolution equations.  It is important to emphasize that there are other groups which also propose  parametrizations for the nuclear effects in the parton distributions  \cite{hkn,deflorian,cteq_nuclear}. Basically, these groups differ in the form of the parametrizations at the initial scale, in the use of different sets of experimental data, in the different nucleon parton densities used in the 
analysis, in the treatment of isospin effects and in the use of sum rules as additional constraints for the evolution.  By construction, these parametrizations describe the current nuclear DIS data. However, the resulting parton distribution sets are very distinct. In particular, the predictions of the different groups for $R_g \equiv xg_A / A.xg_p$ differ largely about the magnitude of the shadowing and the presence or not of the antishadowing (See, e.g. \cite{armesto_review}). Another aspect which is important to emphasize is that the simultaneous description of the neutrino - nucleus  and other nuclear data in terms of a single set  of nuclear parton distribution functions  have being largely discussed in the last years \cite{deflorian,cteq_nuclear,cteq_neu,sal_pau1}. While in Refs. \cite{cteq_nuclear,cteq_neu} the authors have pointed out that the simultaneous description is not feasible, which could be interpreted as a violation of the universality of the parton distributions, in Refs. \cite{sal_pau1,deflorian} the authors demonstrated that if the overall normalization of the experimental data in neutrino DIS is taken into account the current neutrino - nucleus data can also be described in terms of the existing  nPDFs. The latter conclusion have been reinforced recently in the studies presented in Ref. \cite{sal_pau2}. In our study we  assume that the nPDFs are process independent and consider the EPS09 parametrization as a representative model of the nuclear effects.

In Fig. \ref{Ratio-nucle} (a) we present our predictions for the energy dependence of the neutrino - nucleus cross section considering different nuclei. In order to quantify the nuclear effects, in Fig. \ref{Ratio-nucle} (b) we present the nuclear ratio $R$ defined by
\begin{eqnarray}
R \equiv \frac{\sigma_{\nu A}}{A\sigma_{\nu N}}~,
\label{ratio}
\end{eqnarray}
which is equal to one in the absence of effects. As expected, the magnitude of the nuclear effects increases with $A$ and $E_{\nu}$, which is directly associated to the shadowing effect. In particular, we predict a reduction of $\approx 23 \%$ for large energies and $A = 208$. For $A = 16$, which is typical in the Cherenkov neutrino detectors such as IceCube, we predict that the nuclear cross section will be reduced by $\approx 8 \%$ for $E_{\nu}=10^{13}$GeV. In contrast, in the kinematical range of low energies ($E_{\nu} \approx 10^{5}$GeV)  and for light nuclei we predict an enhancement of the nuclear cross section due to the antishadowing effect.
Similar conclusions have been obtained previously in Refs. \cite{varios,zas_nuclear}.

\begin{figure}[t]
\includegraphics*[scale=0.3]{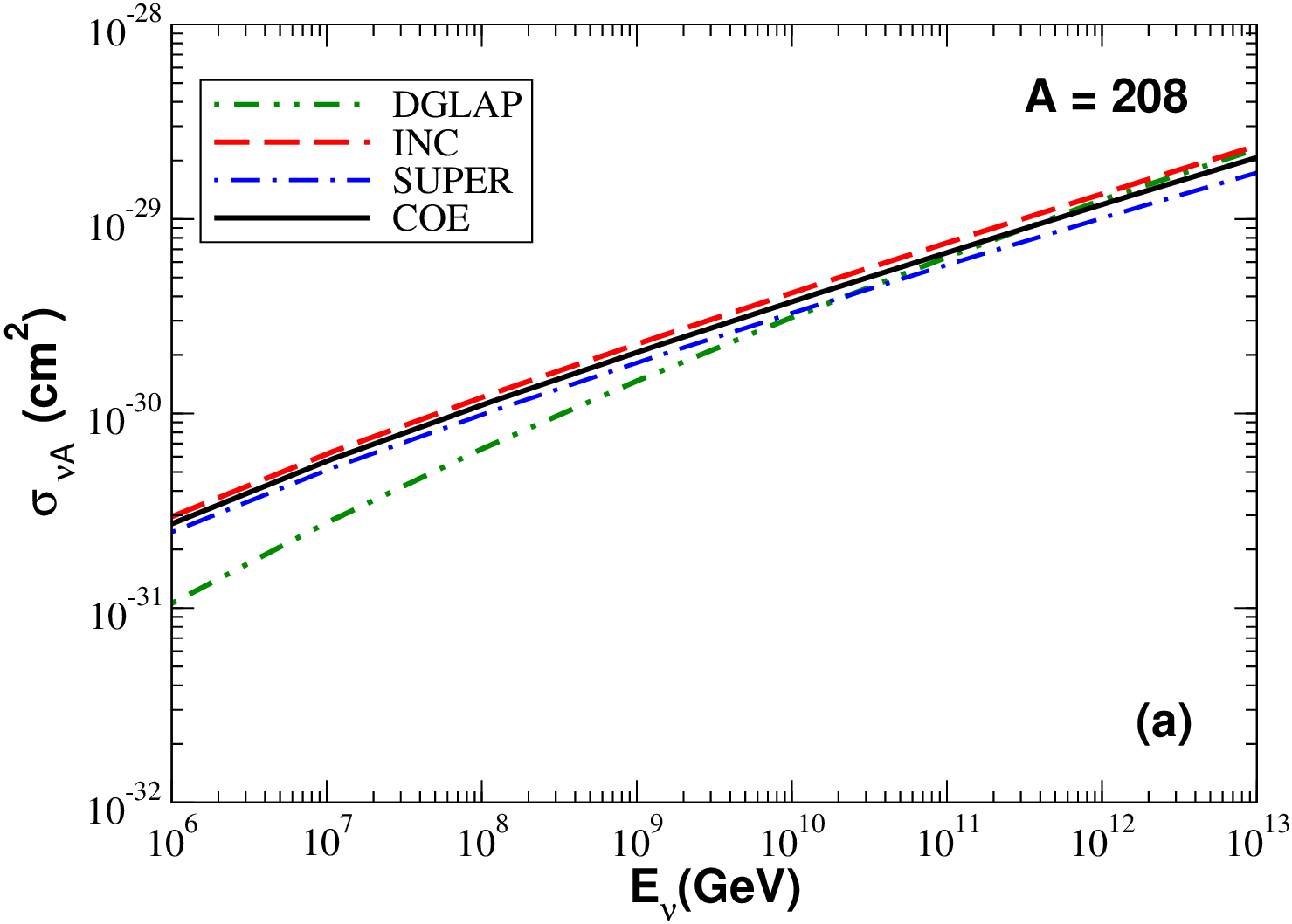}
\includegraphics*[scale=0.3]{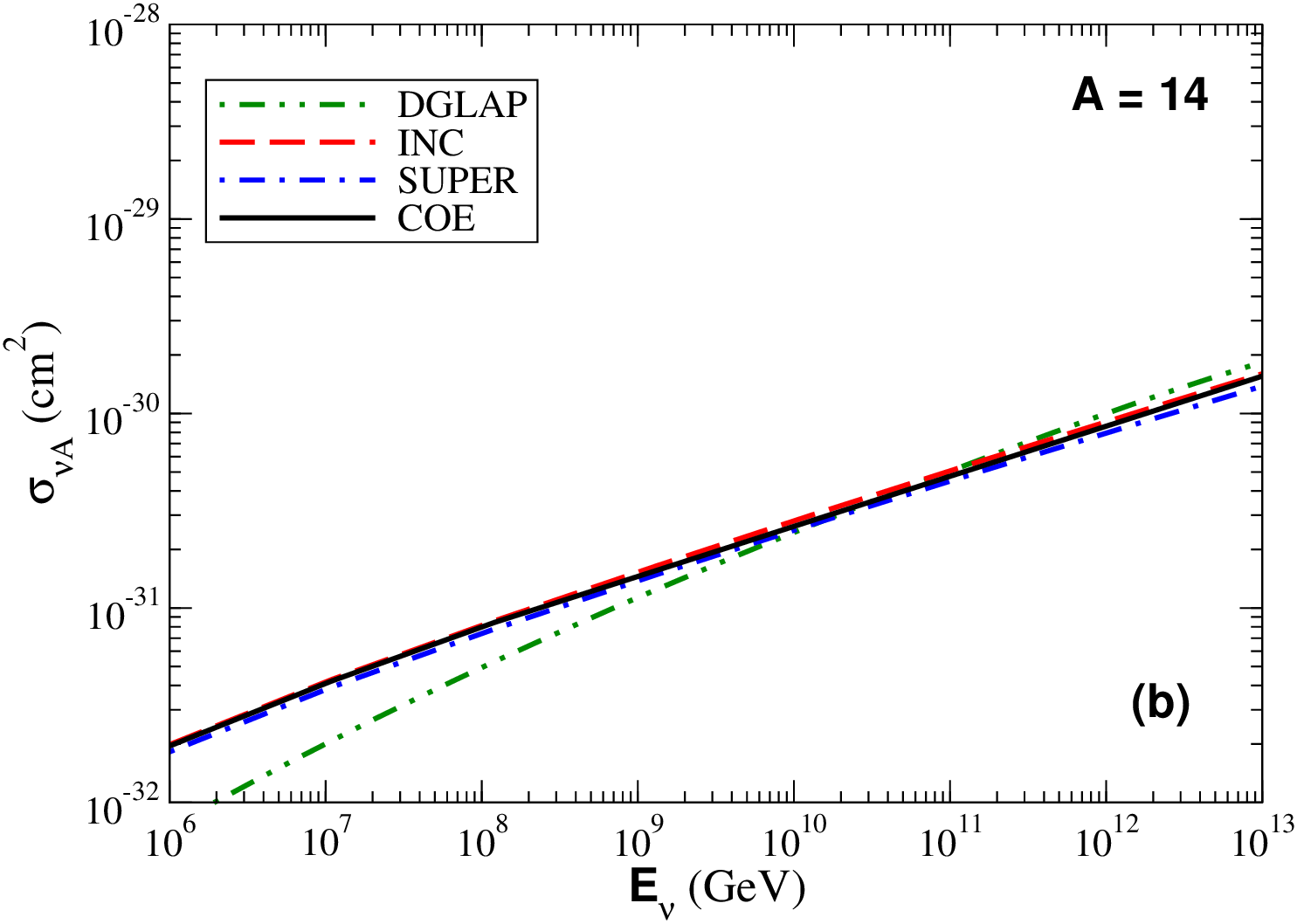}
\caption{(Color online) Energy dependence of the neutrino - nucleus cross section for (a) $A = 208$ and (b) $A = 14$.}
\label{fig:sigma}
\end{figure}

Lets now discuss the non-linear scenario for  the treatment of nuclear effects in the neutrino - nucleus cross section at high energies. The main input in the calculations using the color dipole approach is the  forward dipole - nucleus scattering amplitude ${\cal{N}}^A$ or, equivalently, the dipole-nucleus cross section $\sigma^{dA}$ [See Eq. (\ref{dipnuc})] . In Ref. \cite{armesto_epjc} the author demonstrated that the scarce nuclear data for the nuclear structure function, $F_2^A$, can be described assuming that 
\begin{eqnarray}
{\cal{N}}^A(x,\rr,\rb) = 1 - \exp \left[-\frac{1}{2}A \,T_A(\rb) \, \sigma_{dp}(x,\rr^2)\right] \,\,,
\label{enenuc}
\end{eqnarray}
where $T_A(\rb)$ is the nuclear profile function, which is 
obtained from a 3-parameter Fermi distribution for the nuclear
density normalized to unity and $ \sigma^{dp}$ is the dipole - proton cross section.  More recently, in Refs.  \cite{cazaroto_plb,babi_recent}, we  demonstrated that the current data are not sensitive to the choice of  $\sigma^{dp}$, being equally described using the GBW model and the solution of the rcBK equation. 
The above equation, based on the Glauber-Gribov formalism \cite{gribov},  sums up all the multiple elastic rescattering diagrams of the $q \overline{q}$ pair
and is justified for large coherence length, where the transverse separation $r$ of partons in the multiparton Fock state of the photon becomes a conserved quantity, {\it i. e.} the size of the pair $r$ becomes eigenvalue
of the scattering matrix. We  denote by COE our predictions for $\sigma_{\nu A}$ using Eq. (\ref{enenuc}) and the GBW model as input. As eq. (\ref{enenuc}) represents the classical limit of the Color Glass Condensate  \cite{raju_acta}, it is  expected to be modified by quantum corrections at larger energies than those probed by the current lepton - nucleus  data. The description of   ${\cal{N}}^A$ in the CGC formalism considering these corrections still is an open question. 
We also will consider the following alternative models: (a) The incoherent model (denoted INC hereafter): $\sigma^{dA} = A. \sigma^{dp}$, with $\sigma^{dp}$ given by the GBW model. In this case we are disregarding the nuclear effects. (b) The supersaturated model (denoted SUPER hereafter):  
\begin{eqnarray}
\sigma^{dA}(\rr,x)=\sigma_{0,A}\left[1-\exp\left(-\frac{\rr^{2}Q^{2}_{s,A}(x)}{4}    \right)  \right]
\end{eqnarray}
where  $ \sigma_{0,A}=A^{2/3}\sigma_{0}$ and $Q^{2}_{s,A}=A^{1/3}Q^{2}_{s,p}$. In this model we are assuming that the nucleus is so dense that can be viewed as a large hadron with a continuous particle distribution (For details see Ref. \cite{simone1}). Consequently, we can consider it as a first approximation for  the asymptotic regime of the saturation physics at very large energies. It is important to emphasize that the incoherent and supersaturated models does not describe the current experimental data and should be considered as extreme approaches, which are used in our calculations in order to estimate the  uncertainty associated to the choice of dipole - nucleus cross section.

 In Fig.~\ref{fig:sigma} (a) and (b)  we present our predictions for the neutrino - nucleus cross section for different nuclei considering the linear and non-linear approaches discussed above. We observe a reasonable difference between the DGLAP and non-linear predictions at low energies, which is directly associated with the extrapolation of the dipole model at large - $x$ (More details in the next paragraph). In contrast, we observe that the distinct predictions are similar for ultra - high energies. For $A=208$ and $E_{\nu}=10^{13}$GeV  the DGLAP and INC predictions are identical and can be considered an upper bound for the neutrino - nucleus cross section. On the other hand, the predictions associated to the supersaturated model are $\approx 20 \%$ smaller and can be considered a lower bound.
 In the case of a light nucleus ($A = 14$), the COE and INC predictions are almost identical, which is theoretically expected. In comparison, the predictions of the supersaturated model are $\approx 10 \%$ smaller at $E_{\nu}=10^{13}$GeV. These results indicate that in the nuclear case, differently of the conclusion obtained in Ref. \cite{HEPP} for the neutrino - nucleon cross section,  the theoretical uncertainty in the ultra-high energy regime of the neutrino - nucleus cross section is small and that the use in the calculations of the DGLAP approach corrected by shadowing effects  can be considered a reasonable approximation. 
 
 \begin{figure}[t]
\includegraphics*[scale=0.3]{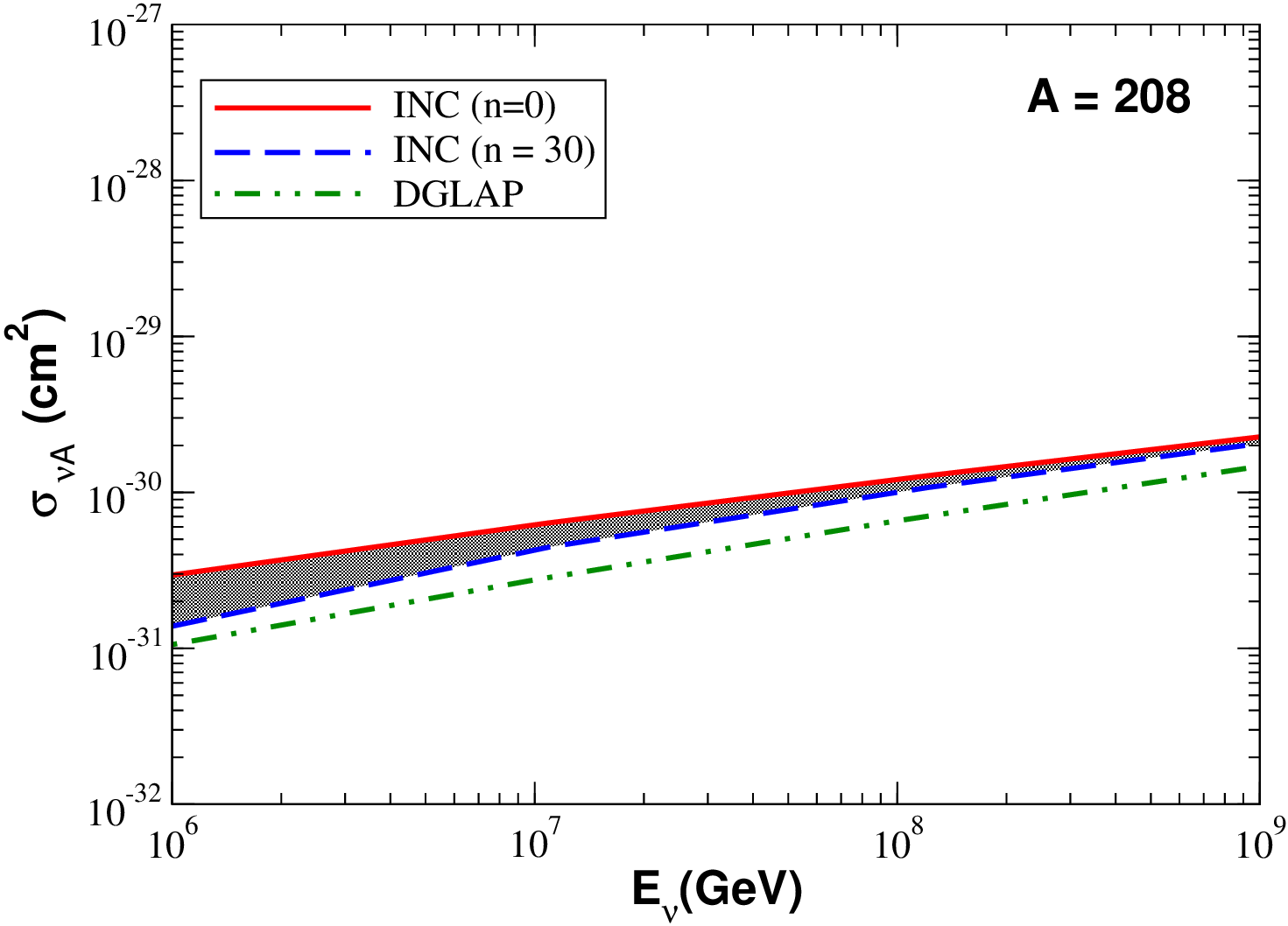}
\caption{(Color online) Comparison between the DGLAP and INC predictions considering different values for the power $n$.}
\label{enes}
\end{figure}
 
Some comments are in order. It is important to emphasize that the dipole approach is only valid at small-$x$ ($\le 10^{-2}$) and that, following previous studies  \cite{gbw,alba_marquet}, we have approximated   the behaviour of the dipole cross section at larger values of $x$  by a multiplicative factor $(1-x)^n$, with $n = 7$, in the calculations. As the total neutrino - nucleus cross section is obtained by an integration on the Bjorken-$x$ [	See Eq. (\ref{total})], we should evaluate the dependence of our predictions in this approximate model for large - $x$ physics. In  Fig. \ref{enes} we compare the DGLAP predictions with those of the incoherent model considering different values of $n$. Similar results are obtained if we consider the other two models for the dipole - nucleus cross section. 
At high energies ($E_{\nu} \ge 10^9)$ the main contribution for the cross section comes from the region of small values of $x$ and the choice of $n$ is not important. On the other hand, at  lower energies, the dipole predictions becomes strongly dependent on the choice of  $n$. We observe that increasing the value of $n$ the INC predictions become closer than the DGLAP one at $E_{\nu} = 10^6$, which is the correct behaviour at low energies.  However, this naive model for the large-$x$ physics cannot be taken seriously, since it does not  contain the scaling violations present in the  DGLAP evolution. The main point is that our predictions at large energies are not dependent on $n$, and our conclusions are robust. Another comment is that we estimated the average inelasticity considering neutrino - nucleus interactions and the linear and non-linear approaches discussed in this Section.  The values found are similar those shown in Fig. \ref{0123} (b) in agreement with the results presented in Ref. \cite{zas_nuclear}, which has demonstrated that the contribution of the nuclear effects in $\langle y \rangle$ is small. Finally, as the initial condition for the linear DGLAP analysis is constrained by the current experimental data, with the extrapolation at low-$x$ being an educated guess,  it is unclear how much of non-linear physics could be hide in the initial condition. Consequently, 
  the agreement of the linear and non-linear predictions at high energies may be, in this sense, accidental.

\section{Summary}
\label{conc}
The ultra-high neutrino-hadron cross-section is a crucial ingredient to the new astrophysics era opened by the recent IceCube measurements. In this paper we
estimate the influence of the  high  partonic density present at high energies and large nuclei in the average inelasticity and in the neutrino - nucleus cross section. We compare the linear DGLAP predictions with those associated to the saturation physics, described by the Color Glass Condensate formalism. The main result of   the present analysis is an uncertainty band for the behaviour of these observables at ultra-high energies , which complement the analysis performed in Ref. \cite{HEPP}.  Our results demonstrated that the saturation effects reduces the average inelasticity and that the predictions of the distinct approaches for the neutrino - nucleus cross section at ultra-high energies are similar.

\begin{acknowledgments}
This work was  financed by the Brazilian funding
agencies CAPES, CNPq and FAPERGS.
\end{acknowledgments}

\end{document}